\documentclass[acmtog]{acmart}

\usepackage{color}
\usepackage{graphicx}
\usepackage{hyperref}
\usepackage{siunitx}
\usepackage{wrapfig}
\usepackage{enumitem}
\usepackage{float}
\usepackage[nolist,nohyperlinks]{acronym}
\newacro{AMD}{age-related macular degeneration}
\newacro{AR}{augmented reality}
\newacro{CPU}{central processing unit}
\newacro{FDR}{false discovery rate}
\newacro{FLORA}{Functional Low-Vision Observer Rated Assessment}
\newacro{FOV}{field of view}
\newacro{GPU}{graphics processing unit}
\newacro{HCI}{human-computer interaction}
\newacro{HMD}{head-mounted display}
\newacro{NFB}{nerve fiber bundle}
\newacro{RGC}{retinal ganglion cell}
\newacro{ROI}{region of interest}
\newacro{RP}{retinitis pigmentosa}
\newacro{SEM}{standard error of the mean}
\newacro{SPV}{simulated prosthetic vision}
\newacro{VR}{virtual reality}
\newacro{XR}{extended reality}
\newacro{VR-SPV}{Virtual Reality Simulated Prosthetic Vision}

\AtBeginDocument{%
  \providecommand\BibTeX{{%
    \normalfont B\kern-0.5em{\scshape i\kern-0.25em b}\kern-0.8em\TeX}}}

\copyrightyear{2022}
\acmYear{2022}
\setcopyright{none}
\acmConference[AHs 2022]{Augmented Humans 2022}{March , 2022}{Preprint}
\acmBooktitle{Augmented Humans 2022 (AHs 2022), March 13--15, 2022, Kashiwa, Chiba, Japan}
\acmDOI{10.1145/3519391.3522752}
\acmISBN{PRE-PRINT, PLEASE CITE DOI:}

\makeatletter
\def\@copyrightspace{\relax}
\makeatother

\begin{document}

\title{Immersive Virtual Reality Simulations of Bionic Vision}

\author{Justin Kasowski}
\affiliation{
    \institution{University of California, Santa Barbara}
    \city{Santa Barbara}
    \state{CA}
    \country{USA}
}
\email{justin_kasowski@ucsb.edu}

\author{Michael Beyeler}
\affiliation{
    \institution{University of California, Santa Barbara}
    \city{Santa Barbara}
    \state{CA}
    \country{USA}
}
\email{mbeyeler@ucsb.edu}

\renewcommand{\shortauthors}{Kasowski \& Beyeler}

\begin{abstract}

Bionic vision uses neuroprostheses to restore useful vision to people living with incurable blindness. 
However, a major outstanding challenge is predicting what people ``see'' when they use their devices.
The limited field of view of current devices necessitates head movements to scan the scene, which is difficult to simulate on a computer screen. In
addition, many computational models of bionic vision lack biological realism.
To address these challenges, we present VR-SPV, an open-source virtual reality toolbox for simulated prosthetic vision that uses a psychophysically validated computational model to allow sighted participants to ``see through the eyes'' of a bionic eye user.
To demonstrate its utility, we systematically evaluated how clinically reported visual distortions affect performance in a letter recognition and an immersive obstacle avoidance task.
Our results highlight the importance of using an appropriate phosphene model when predicting visual outcomes for bionic vision.

\end{abstract}



\begin{teaserfigure}
  \includegraphics[width=\textwidth]{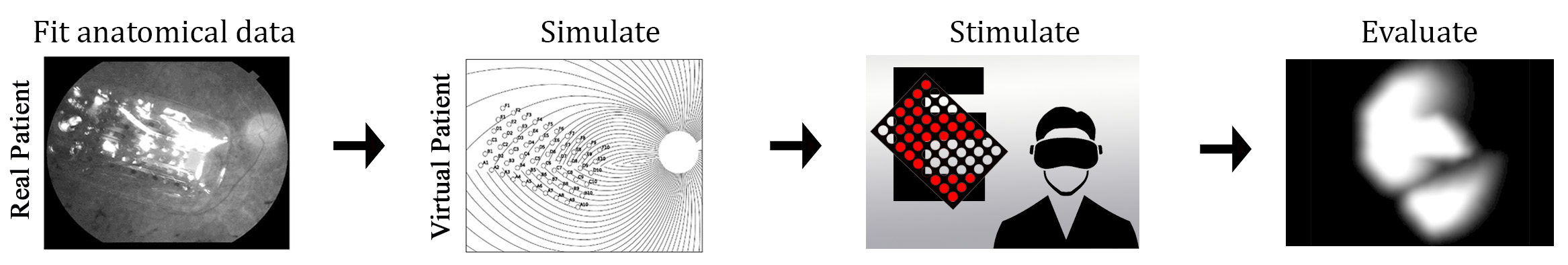}
  \caption{Immersive virtual reality simulations of bionic vision. A microelectrode array is implanted in the eye to stimulate the retina $\rightarrow$ Anatomical data is used to position a simulated electrode array on a simulated retina to create a ``virtual patient'' $\rightarrow$ Visual input from a virtual reality environment acts as stimulus for the simulated implant to generate a realistic prediction of \acf{SPV} $\rightarrow$ The rendered \acs{SPV} image is presented to the virtual patient and behavioral metrics are recorded }
  \Description{Figure 1: Left Panel. An image labelled as ``Fit anatomical data, real patient" showing a fundus image of an implanted retinal prosthesis.  An arrow points to the next panel labelled ``Simulate, virtual patient'' which shows a simulated electrode array overtop of a retinal map drawing. An arrow points to the next panel which is labelled ``Stimulate'' and shows a person with a virtual reality headset and  the Letter E with a simulated electrode array over it.  An arrow points to the final panel which is labelled “Evaluate” and shows a distorted version of the parts of the E that activated the simulated electrode array}
  \label{fig:teaser}
\end{teaserfigure}

\maketitle

\section{Introduction}

The World Health Organization has estimated that by the year 2050, roughly 114.6 million people will be living with incurable blindness and 587.6 million people will be affected by severe visual impairment \cite{bourne_magnitude_2017}.
Although some affected individuals can be treated with surgery or medication, there are no effective treatments for many people blinded by severe degeneration or damage to the retina, the optic nerve, or cortex. In such cases, an electronic visual prosthesis (``bionic eye'') may be the only option \cite{fernandez_development_2018}.
Analogous to cochlear implants, these devices electrically stimulate surviving cells in the visual pathway to evoke visual percepts (``phosphenes''). 
The phosphenes produced by current prostheses generally provide users with an improved ability to localize high-contrast objects and perform basic orientation \& mobility tasks \cite{ayton_update_2020}, but are not yet able to match the acuity of natural vision.

Despite their potential to restore vision to people living with incurable blindness, the number of bionic eye users in the world is still relatively small (roughly $500$ retinal prostheses implanted to date).
To investigate functional recovery and experiment with different implant designs, researchers have therefore been developing \ac{VR} prototypes that rely on \acf{SPV}.
The classical method relies on sighted subjects wearing a \ac{VR} headset, who are then deprived of natural viewing and only perceive phosphenes displayed in a \ac{HMD}. 
This allows sighted participants to ``see through the eyes'' of the bionic eye user, taking into account their head and/or eye movements as they explore a virtual environment \cite{kasowski_towards_2021}.

However, most \ac{SPV} studies simply present stimuli  on a computer monitor or an \ac{HMD}, without taking into account eye movements, head motion, or locomotion \cite{kasowski_furthering_2021}. This leads to a low level of immersion \cite{kardong-edgren_call_2019, nijholt_immersion_2009}, which refers to 
technical manipulations that separate the existence of the physical world from the virtual world \cite{miller_level_2016}. Seeing the real world, using a keyboard or joystick to move, and sounds present in the real environment are factors which lead to lower levels of immersion.
However, the role of immersion for behavioral tasks in \ac{SPV} is still unclear as no previous study has assessed whether behavioral performance is comparable across monitor-based and \ac{HMD}-based versions of the same task.

In addition, most current prostheses provide a very limited \ac{FOV}; for example, the artificial vision generated by Argus II~\cite{luo_argus_2016}, the most widely adopted retinal implant thus far, is restricted to roughly $10 \times 20$ degrees of visual angle. This forces users to scan the environment with strategic head movements while attempting to piece together the information \cite{erickson-davis_what_2021}. The emergence of immersive \ac{VR} as a research tool provides researchers with the ability to simulate this in a meaningful way. 

Moreover, a recent literature review found that most \ac{SPV} studies relied on phosphene models with a low level of biological realism \cite{kasowski_furthering_2021}.
It is therefore unclear how the findings of most \ac{SPV} studies would translate to real bionic eye users.

To address these challenges, we make three contributions:

\begin{itemize}[topsep=0pt]
    \item We present VR-SPV, a \acf{VR} reality toolbox for \acf{SPV} that allows sighted participants to ``see through the eyes'' of a bionic eye user.
    \item Importantly, we use an established and psychophysically validated computational model of bionic vision \cite{beyeler_pulse2percept_2017} to generate realistic \ac{SPV} predictions.
    \item We systematically evaluate how different display types (\ac{HMD} or monitor) affect behavioral performance in a letter recognition and an obstacle avoidance task. To the best of our knowledge, this is the first \ac{SPV} study that uses a within-subjects design to allow for a direct comparison between display types of the same tasks.

\end{itemize}

\section{Background}
\label{sec:background}

\subsection{Restoring Vision with a Bionic Eye}
The only FDA-approved technology for the treatment of retinal degenerative blindness are visual neuroprostheses. These devices consist of an electrode array implanted into the eye or brain that is used to artificially stimulate surviving cells in the visual system. 
Two retinal implants are already commercially available (Argus II: 60 electrodes, Second Sight Medical Products, Inc. \cite{luo_argus_2016}; Alpha-AMS (1600 electrodes, Retina Implant AG,
\cite{stingl_artificial_2013}), and many other emerging devices have reached the clinical or pre-clinical stage \cite{lorach_photovoltaic_2015, ferlauto_design_2018, ayton_first--human_2014}. 
In order to create meaningful progress within these devices, there is a growing need to understand how the vision these devices provide differs from natural sight \cite{erickson-davis_what_2021}. 

One common fallacy is the assumption that each electrode produces a focal spot of light in the visual field \cite{perez-yus_depth_2017, sanchez-garcia_indoor_2019, kvansakul_sensory_2020}. This is known as the scoreboard model, which implies that creating a complex visual scene can be accomplished simply by using the right combination of pixels (analogous to creating numbers on a sports stadium scoreboard).
On the contrary, recent work suggests that phosphenes vary in shape and size, differing considerably across subjects and electrodes \cite{fine_pulse_2015,luo_long-term_2016, beyeler_model_2019}. 
 
\begin{figure}[!b]
\centering
\includegraphics[width=\columnwidth]{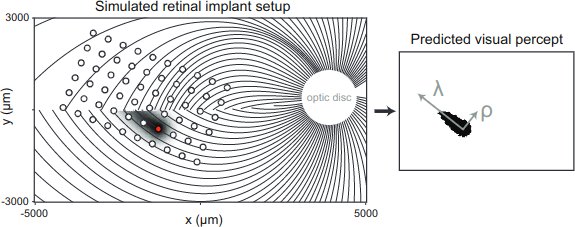}
\caption{A simulated map of retinal \acsp{NFB} (\emph{left})
    can account for visual percepts (\emph{right}) 
    elicited by retinal implants (reprinted with permission from \cite{beyeler_model-based_2019}).
    \textbf{\emph{Left}:} Electrical stimulation (red circle) of a \acs{NFB} (black lines) could activate retinal ganglion cell bodies peripheral to the point of stimulation,
    leading to tissue activation (black shaded region) elongated
    along the \acs{NFB} trajectory away from the optic disc (white circle).
    \textbf{\emph{Right}:} The resulting visual percept appears elongated; its shape can be described by two parameters, $\lambda$ and 
    $\rho$.
}
\Description{Figure 2: Left panel: Simulated retinal implant setup,  consisting of an electrode array situation on top of a drawing of a retina map.  A single electrode is red and the percept it generates is a black streak around the electrode.  Right panel: Predicted Visual Percept, a panel that has isolated the percept and describes it by its elongation, labelled lambda, and its spread, labelled rho}
\label{fig:axonmap}
\end{figure}

Increasing evidence suggests that perceived phosphene shape in an epiretinal implant is a result of unintended stimulation of \acp{NFB} in the retina \cite{rizzo_perceptual_2003,beyeler_model_2019}. These \acp{NFB} follow polar trajectories~\cite{jansonius_mathematical_2012} away from the horizontal meridian, forming arch-like projections into the optic nerve (Fig.~\ref{fig:axonmap}, \emph{left}).
Stimulating a \ac{NFB} would result in the activation of nearby \acp{RGC} that are upstream in the trajectory, resulting in phosphenes that appear elongated (Fig.~\ref{fig:axonmap}, \emph{right}).

Ref.~\cite{beyeler_model_2019} demonstrated through simulations that the shape of elicited phosphenes closely followed \ac{NFB} trajectories.
Their computational model assumed that an axon's sensitivity to electrical stimulation:
\begin{enumerate}[topsep=0pt,itemsep=-1ex,partopsep=0pt,parsep=1ex,leftmargin=24pt,label=\roman*.]
    \item decayed exponentially with $\rho$ as a function of distance from the stimulation site,
    \item decayed exponentially with $\lambda$ as a function of distance from the cell body, measured as axon path length.
\end{enumerate}
In other words, the values of $\rho$ and $\lambda$ in this model dictate the size and elongation of phosphenes, respectively. This may drastically affect visual outcomes, as large values of $\lambda$ are thought to distort phosphene shape.

\subsection{Related Work}
\label{sec:related_work}

The use of virtual reality has emerged as a tool for assisting users with low vision (see \cite{kasowski_furthering_2021} for a review of recent literature). This includes not just accessibility aids, but also simulations aimed at understanding low vision. A number of previous \ac{SPV} studies have focused on assessing
the impact of different stimulus and model parameters (e.g., phosphene size, phosphene spacing, flicker rate) on measures of visual acuity.
Stimuli for these low-level visual function tests were often presented on monitors~\cite{vurro_simulation_2014,lu_estimation_2012} or in \acp{HMD}~\cite{wu_prosthetic_2014, cao_eye-hand_2017}.
Some studies also tested the influence of \ac{FOV}~\cite{sanchez-garcia_influence_2020, thorn_virtual_2020} and eye gaze compensation~\cite{titchener_gaze_2018} on acuity.
Others focused on slightly more complex tasks such as letter~\cite{zhao_reading_2011}, word~\cite{perez_fornos_reading_2011}, face~\cite{denis_human_2013,chang_facial_2012}, and object recognition~\cite{zhao_image_2010, wang_cross-task_2018, mace_simulated_2015}.
In most setups, participants would view \ac{SPV} stimuli in a conventional \ac{VR} \ac{HMD}, but some studies also relied on smart glasses to present \ac{SPV} in \ac{AR}.

However, because most of the studies mentioned above relied on the scoreboard model, 
it is unclear how their findings would translate to real bionic eye users.
Although some studies attempted to address phosphene distortions \cite{wu_prosthetic_2014,guo_enhanced_2014,subramaniam_simulating_2012}, most did not account for the neuroanatomy (e.g., \ac{NFB} trajectories) when deciding how to distort phosphenes. 
Only a handful of studies have incorporated a great amount of neurophysiological detail into their setup~\cite{josh_psychophysics_2013,vurro_simulation_2014,wang_cross-task_2018,thorn_virtual_2020}, only two of which~\cite{wang_cross-task_2018,thorn_virtual_2020} relied on an established and psychophysically validated model of \ac{SPV}.
One notable example is the study by Thorn \emph{et al.}~\cite{thorn_virtual_2020}, which accounted for unintentional stimulation of axon fibers in the retina by adding a fixed ``tail'' length to each phosphene. However, a fixed-length tail is a simplification of the model \cite{beyeler_model_2019} as the size of phosphenes (and their tails) have been shown to vary with stimulation parameters such as amplitude, frequency, and pulse duration \cite{nanduri_frequency_2012}. 

In addition, being able to move around as one would in real life has shown to significantly increase the amount of immersion a user experiences \cite{nijholt_immersion_2009}.
However, the level of immersion offered by most \ac{SPV} studies is relatively low, as stimuli are often presented on a screen \cite{ying_recognition_2018, wang_cross-task_2018}.
In contrast, most current prostheses provide a very limited \ac{FOV} (e.g., Argus II: $10 \times 20$ degrees of visual angle), which requires users to scan the environment with strategic head movements while trying to piece together the information \cite{erickson-davis_what_2021}.
Furthermore, Argus II does not take into account the eye movements of the user when updating the visual scene, which can be disorienting for the user.
Ignoring these \ac{HCI} aspects of bionic vision can result in unrealistic predictions of prosthetic performance, sometimes even exceeding theoretical acuity limits (as pointed out by \cite{caspi_assessing_2015}).
 
In summary, previous \ac{SPV} research has assumed that the scoreboard model produces phosphenes that are perceptually similar to real bionic vision \cite{vergnieux_wayfinding_2014, zhao_reading_2011},
and that findings from an \ac{HMD}-based task would more accurately represent the experience of a bionic eye user than a monitor version \cite{zapf_towards_2014, sanchez-garcia_influence_2020, thorn_virtual_2020}. 
In this paper we aim to systematically evaluate these assumptions with a within-subjects (repeated measures) design, allowing for direct comparisons in performance across different model parameters and display conditions. 

\section{Methods}

\subsection{VR-SPV: A Virtual Reality Toolbox for Simulated Prosthetic Vision}

The VR-SPV system consisted of either a wireless head-mounted \ac{VR} headset (HTC VIVE Pro Eye with wireless adapter, HTC Corporation) or a standard computer monitor (Asus VG248QE, 27in, 144Hz, 1920x1080p). Both \ac{HMD} and monitor versions used the same computer for image processing (Intel i9-9900k processor and an Nvidia RTX 2070 Super GPU with 16GB of DDR4 memory).
All software was developed using the Unity development platform, consisting of a combination of C\# code processed by the \ac{CPU} and fragment/compute shaders processed by the \ac{GPU}.
The entire software package, along with a more in-depth explanation, is available at \url{https://github.com/bionicvisionlab/BionicVisionXR}.

The workflow for simulating bionic vision was as follows:

\begin{enumerate}[topsep=0em,itemsep=-1ex,partopsep=0em,parsep=1ex,leftmargin=6ex,label=\roman*.]
    \item Image acquisition: Utilize Unity's virtual camera to acquire the scene at roughly 90 frames per second and downscale to a target texture of $86 \times 86$ pixels.
    \item Image processing: Conduct any image preprocessing specified by the user. Examples include grayscaling, extracting and enhancing edges, and contrast maximization. In the current study, the image was converted to grayscale and edges were extracted in the target texture with a $3 \times 3$ Sobel operator. 
    \item Electrode activation: Determine electrode activation based on the visual input as well as the placement of the simulated retinal implant. In the current study, a $3 \times 3$ Gaussian blur was applied to the preprocessed image to average the grayscale values around each electrode's location in the visual field. This gray level was then interpreted as a current amplitude delivered to a particular electrode in the array.
    \item Phosphene model: Use Unity shaders to convert electrode activation information to a visual scene in real time. The current study re-implemented the axon map model available in \emph{pulse2percept} \cite{beyeler_pulse2percept_2017} using shaders. 
    \item Phosphene rendering: Render the elicited phosphenes either on the computer monitor or the \ac{HMD} of the \ac{VR} system.
\end{enumerate}

The VR-SPV system is designed to handle any retinal implant by allowing users to specify the location and size of each electrode in the simulated device. It can also handle other phosphene models, including cortical models, by replacing the model provided by \emph{pulse2percept} with any phosphene model of their choosing.

While not considered in this study, VR-SPV can also be used to model temporal interactions by integrating electrode activation from previous frames or by only rendering at a specific frequency. 
The software is also capable of utilizing the VIVE's eye tracking hardware to elicit a ``gaze lock''. 
This function moves the rendered image to the center of the user's gaze, attempting to replicate the inability of a prosthetic user to scan the presented image with eye movements. Neither of these optional functions were used in this study as they were not the focus of the current work, and it was unclear how these settings would influence any findings on the parameters being studied.
VR-SPV also includes a function to change the source of the visual input from a virtual environment to the \ac{HMD}'s front-facing camera for supporting \ac{AR} applications.

\begin{figure*}[t]
    \centering
    \includegraphics[width=\linewidth]{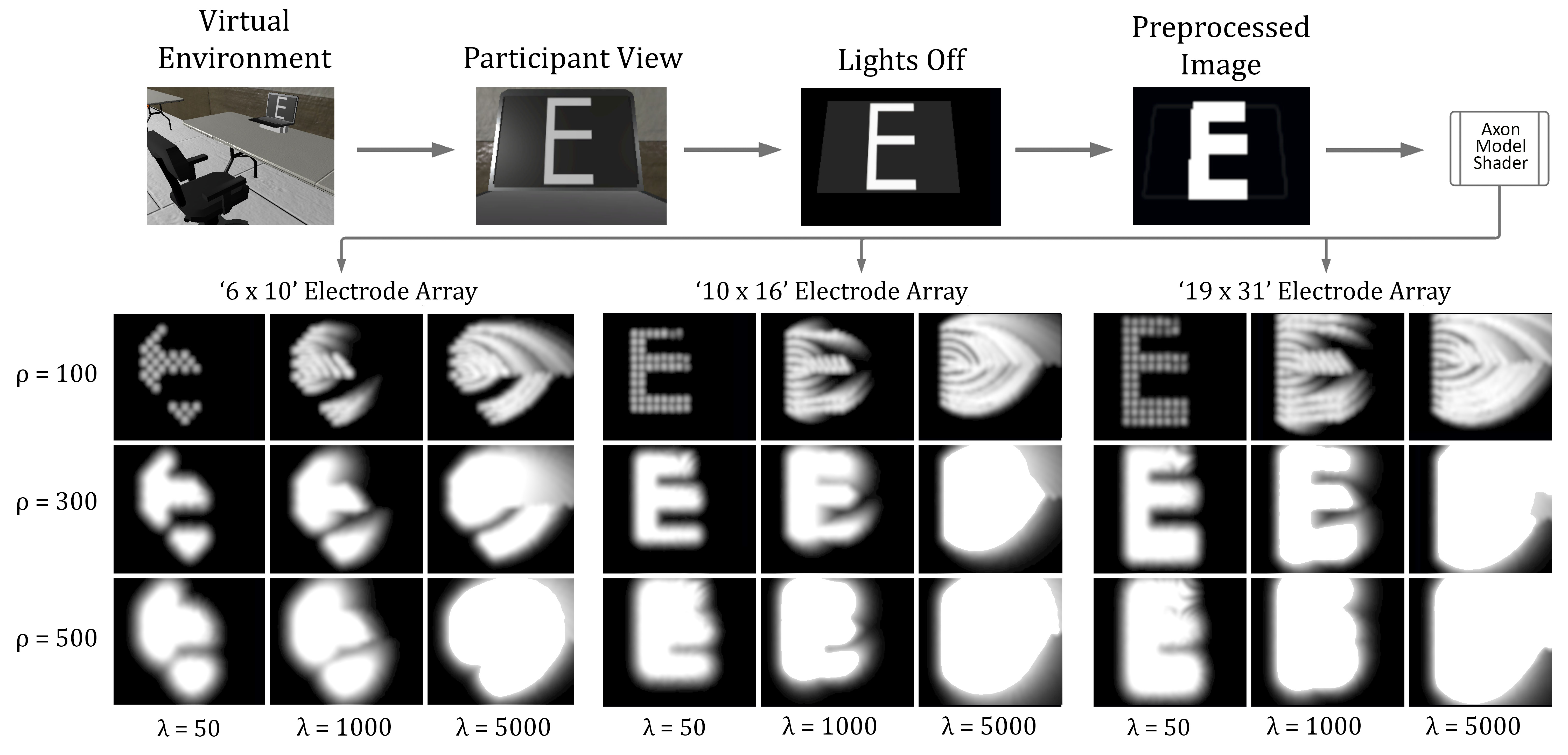}
    \caption{Letter recognition task. \textbf{\emph{Top:}} The lights in the virtual room are turned off and the image seen by the user is passed to the preprocessing shader which performs edge extraction/enhancement before the axon model shader renders \ac{SPV}. Modeled after \cite{cruz_argus_2013}.
    \textbf{\emph{Bottom:}} Output of the axon model shader across the various devices and $\rho$ / $\lambda$ combinations.}
    \Description{Figure 3: A picture of a chair and laptop labelled Virtual Environment. An arrow points to a zoomed in picture of the laptop screen with the letter ``E'' labelled Participant View. An arrow points to a similar picture but only the laptop screen is visible, labelled Lights Off.  An arrow points to a processed image that is a thicker version of the letter ``E'' from the previous panel, labelled Preprocessed Image.  An arrow points to a small box with the words Axon Model Shader.  The axon Model shader points to three different panels showing the outputs of simulated prosthetic vision across different devices for the preprocessed image }
    \label{fig:exp-letter-recognition}
\end{figure*}

\subsection{Simulated Prosthetic Vision}

The underlying phosphene model for this experiment was a re-implementation of the pyschophysically validated axon map model \cite{beyeler_model_2019} provided by \emph{pulse2percept} \cite{beyeler_pulse2percept_2017}.
To support real-time execution, an initial mapping of each electrode's effects on the scene were pre-calculated with \emph{pulse2percept} before starting the experiment.
The shape of the elicited phosphenes was based on the retinal location of the simulated implant as well as model parameters $\rho$ and $\lambda$ (see Section~\ref{sec:background}). As can be seen in Fig.~\ref{fig:axonmap} (\emph{left}), electrodes near the horizontal meridian 
activated cells close to the end of the \acp{NFB}, limiting the potential of elongation along an axon. This resulted in more circular phosphenes, whereas other electrodes were predicted to produce elongated percepts that differed in angle based on whether they fell above or below the horizontal meridian.

We were particularly interested in assessing how different \ac{SPV} model parameters affected behavioral performance.
Importantly, $\rho$ and $\lambda$ vary drastically across patients \cite{beyeler_model_2019}.
Although the reason for this is not fully understood, it is clear that the choice of these parameter values may drastically affect the quality of the generated visual experience.
To cover a broad range of potential visual outcomes, we simulated nine different conditions by combining $\rho=\{100, 300, 500\}$ \SI{}{\micro\meter} with $\lambda=\{50, 1000, 5000\}$\SI{}{\micro\meter}.

We were also interested in how the number of electrodes in an implant and the associated change in \ac{FOV} affected behavioral performance.
In addition to simulating Argus II, we created two hypothetical near-future devices that used the same aspect ratio and electrode spacing, but featured a much larger number of electrodes.
Thus the three devices tested were:
\begin{itemize}[topsep=0pt]
    \item Argus II: $6 \times 10 = 60$ equally spaced electrodes situated \SI{575}{\micro\meter} apart in a rectangular grid. To match the implantation strategy of Argus II, the device was simulated at \SI{-45}{\degree} with respect to the horizontal meridian in the dominant eye.
    \item Argus III (hypothetical): $10 \times 16 = 160$ electrodes spaced \SI{575}{\micro\meter} apart in a rectangular grid implanted at \SI{0}{\degree}. A recent modeling study suggests that this implantation angle might minimize phosphene streaks \cite{beyeler_model-based_2019}.
    \item Argus IV (hypothetical): $19 \times 31 = 589$ electrodes spaced \SI{575}{\micro\meter} apart in a rectangular grid implanted at \SI{0}{\degree}.
\end{itemize}

\subsection{Participants}
We recruited 17 sighted participants (6 female and 11 male; ages $27.4 \pm 5.7$ years) from the student pool at the University of California: Santa Barbara. Participation was voluntary and subjects were informed of their right to freely withdraw for any reason. Recruitment and experimentation followed protocols approved by the university's Institutional Review Board, along with limitations and safety protocols approved by the university's COVID-19 response committee.

None of the participants had previous experience with \ac{SPV}.
Participants were split into two equally sized groups; one starting with the \ac{HMD}-based version of the first experiment while the other started with the monitor-based version. 

In order to get accommodated with the \ac{SPV} setup,
participants began each task with the easiest block; that is, the scoreboard model ($\lambda$=\SI{50}{\micro\meter}) with the smallest possible phosphene size and the highest number of electrodes.  
The order of all subsequent blocks was randomized for each participant. 

\begin{figure*}[t]
    \centering
    \includegraphics[width=\linewidth]{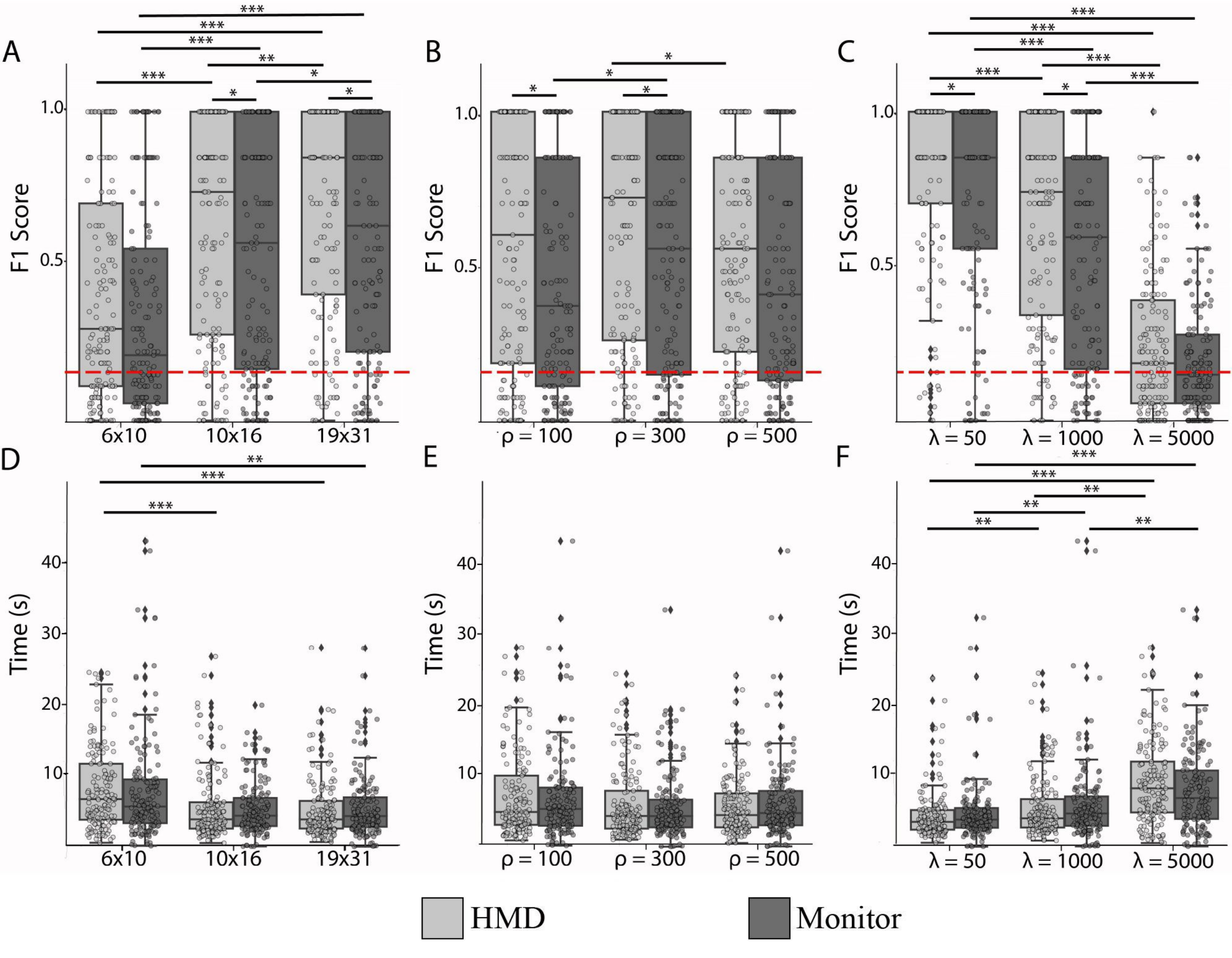}
    \caption{Letter recognition task.  Data points represent each subject's average performance in a block with boxplots displaying median and interquartile ranges.  \textbf{\emph{Top: }}  Average F1 score across blocks for each subject within the condition specified by the x-axis.  \textbf{\emph{Bottom}}: Average time across blocks for each subject within the condition specified by the x-axis. Statistical significance was determined using ART ANOVA (*<.05, **<.01, ***<.001).}
    \Description{Six box plots for the letter recognition task that are fully explained plot by plot in the results}
    \label{fig:results-letter}
\end{figure*}

\section{Experiments and Results}

To study the impact of \ac{SPV} parameters and level of immersion, we replicated two popular tasks from the bionic vision literature.
The first task was a basic letter recognition experiment \cite{cruz_argus_2013}, tasking participants with identifying the letter presented to them. 
The second one was a more immersive orientation \& mobility task, requiring subjects to walk down a virtual hallway while avoiding obstacles \cite{he_improved_2020}.
 
To allow for a direct comparison across all conditions, we chose a within-subjects, randomized block design.
This systematic side-by-side comparison minimized the risk of learning effects and other artifacts that may arise from inhomogeneity between groups, allowing for meaningful statistics with a relatively small number of subjects. 

The procedures and results for each task are presented separately below, followed by a joint discussion on both experiments in the subsequent sections.

\subsection{Task 1: Letter Recognition}

\subsubsection{Original Task}
The first experiment was modeled after a letter recognition task performed by Argus II recipients \cite{cruz_argus_2013}.
In the original task, following a short training period, participants were instructed to identify large and bright white letters presented on a black screen situated \SI{0.3}{\meter} in front of them. 
Participants were given unlimited time to respond.
The experiment was carried out in a darkened room.
Both the initial training period and the actual experiment featured all 26 letters of the alphabet. The letters were grouped by similarity and tested in batches of 8, 8, and 10 letters.

\subsubsection{Experimental Setup and Procedure}

To emulate the experiment described in \cite{cruz_argus_2013}, we
carefully matched our virtual environment to the experimental setup of the original task.
The setup mainly consisted of a virtual laptop on top of a virtual desk (Fig.~\ref{fig:exp-letter-recognition}).
A virtual monitor was positioned \SI{0.3}{\meter} in front of the user's head position.
In agreement with the original task, participants were presented letters that were \SI{22.5}{\centi\meter} tall (subtending \SI{41.112}{\degree} of visual angle)
in True Type Century Gothic font. For the monitor version of the task, the camera was positioned at the origin and participants could simulate head movements by using the mouse.

\begin{table*}[tb]
	\centering
	\begin{tabular}[]{|c|cc|cc|}
		\hline
		& \multicolumn{2}{c|}{F1 Score ($\pm$ Std Dev)} & \multicolumn{2}{c|}{Mean Time (s)  ($\pm$ Std Dev)} \\
 		 & HMD & Monitor & HMD & Monitor  \\
\hline
		 06x10 Array  &  0.411 ($\pm$ 0.341 )  &   0.344 ($\pm$ 0.339 ) &  8.312  ($\pm$ 5.685 )  &   7.979  ($\pm$ 7.367 ) \\

		 10x16 Array  &  0.628 ($\pm$ 0.361 )  &   0.546 ($\pm$ 0.380 ) &  5.853  ($\pm$ 4.808 )  &   \textbf{5.567  ($\pm$ 3.608 )} \\

		 19x31 Array  &  \textbf{0.699 ($\pm$ 0.347 )}  &   \textbf{0.596 ($\pm$ 0.373 )} &  \textbf{5.379  ($\pm$ 4.164 )}  &   5.661  ($\pm$ 4.373 ) \\
		\hline
		 $\rho$=100  &  0.570 ($\pm$ 0.388 )  &   0.467 ($\pm$ 0.381 ) &  7.415  ($\pm$ 5.874 )  &   7.007  ($\pm$ 6.120 ) \\

		 $\rho$=300  &  \textbf{0.620 ($\pm$ 0.366 )}  &   \textbf{0.540 ($\pm$ 0.379 )} &  6.173  ($\pm$ 4.879 )  &   \textbf{5.829  ($\pm$ 4.686 )} \\

		 $\rho$=500  &  0.548 ($\pm$ 0.352 )  &   0.479 ($\pm$ 0.377 ) &  \textbf{5.956  ($\pm$ 4.267 )}  &   6.371  ($\pm$ 5.483 ) \\
		\hline
		 $\lambda$=50  &  \textbf{0.824 ($\pm$ 0.267 )}  &   \textbf{0.750 ($\pm$ 0.329 )} &  \textbf{4.540  ($\pm$ 3.408 )}  &   \textbf{5.034  ($\pm$ 4.403 )} \\

		 $\lambda$=1000  &  0.665 ($\pm$ 0.331 )  &   0.543 ($\pm$ 0.362 ) &  5.698  ($\pm$ 4.338 )  &   6.074  ($\pm$ 5.780 ) \\

		 $\lambda$=5000  &  0.248 ($\pm$ 0.229 )  &   0.193 ($\pm$ 0.188 ) &  9.307  ($\pm$ 5.906 )  &   8.099  ($\pm$ 5.700 ) \\
		\hline
	\end{tabular}
	\caption{Letter recognition task: Average performance and time per trial across conditions. Best performances (highest F1/shortest time) for each grouping are presented in bold.}
	\label{tab:average-letter}
\end{table*}

Each combination of 3 devices $\times$ 3 $\rho$ values $\times$ 3 $\lambda$ values were implemented as a block, resulting in a total of 27 blocks. All 27 blocks were completed twice; once for the \ac{HMD} version of the task, and once for the monitor version of the task.
Rather than presenting all 26 letters of the alphabet (as in the original experiment), we limited our stimuli to the original Snellen letters (C, D, E, F, L, O, P, T, Z) for the sake of feasibility.

All nine Snellen letters were presented in each block, resulting in a total of 243 trials.
Participants were limited to 1 minute per trial, after which the virtual monitor would go dark and the participant had to select a letter before the experiment continued. 

To acclimate participants to the task and controls, we had them perform an initial practice trial using normal vision.
After that, the lights in the virtual room were turned off and the VR-SPV toolbox was used to generate \ac{SPV}.
To mimic the training session of \cite{cruz_argus_2013}, participants completed three practice trials using \ac{SPV} at the beginning of each block.
Participants were able to repeat each practice trial until they had selected the correct letter.
To prevent participants from memorizing letters seen during practice trials, we limited practice trials to the letters Q, I, and N.

Participant responses and time per trial were recorded for the entirety of the experiment.

\subsubsection{Performance Evaluation}
Perceptual performance was assessed using F1 scores, which represent the harmonic mean between precision and recall, allowing for a slight penalty towards false positive choices compared to recall (proportion correct) on its own. 
This had the advantage of eliminating bias towards specific letter choices.
F1 values were calculated for each block using the scikit-learn `f1\_score' function \cite{pedregosa_scikit-learn_2011}. We also measured time per trial with the assumption that easier trials could be completed faster than trials that were more difficult. 

Due to ceiling and floor effects, neither outcome measure (F1 scores and time per trial) were normally distributed, violating the assumptions of the standard ANOVA.
We therefore performed a subsequent aligned rank transform (ART) with the R package ARTool \cite{wobbrock_aligned_2011} for both F1 scores and time per trial. This method of analysis allows for a factorial ANOVA to be performed on repeated measures, non-uniform data, and lower subject counts \cite{wobbrock_aligned_2011}. 
Post-hoc analyses were performed on significant groups by analyzing the rank-transformed contrasts \cite{elkin_aligned_2021}. The Tukey method  \cite{tukey_comparing_1949} was used to adjust $p$-values to correct for multiple comparisons. All code used in the analysis, along with the raw data, is provided at \url{https://github.com/bionicvisionlab/2022-kasowski-immersive}.

\subsubsection{Results}

Results from the letter recognition task are summarized in Table \ref{tab:average-letter} and distributions are plotted in Fig.~\ref{fig:results-letter}. Group F-values, along with their significance, are reported in Table \ref{tab:letter-anova}.
Each data point in Fig.~\ref{fig:results-letter} represents a subject's F1 score (Fig.~\ref{fig:results-letter}A--C) and time per trial (Fig.~\ref{fig:results-letter}D--F) across all letters in a block.
F1 score ranged from 0 to 1 with higher values representing better performance. Assuming a different letter is chosen for each selection, a chance-level F1 score would equal the probability for randomly guessing the correct letter ($\frac{1}{9}=0.1111$).

\begin{table}[b]
\begin{tabular}{|l|cc|cc|}
\hline
\multicolumn{1}{|l|}{}    & \multicolumn{2}{c|}{{\bf F1 Score}}           & \multicolumn{2}{c|}{{\bf Time}} \\
\multicolumn{1}{|l|}{}    & F-Value        & \multicolumn{1}{c|}{Signif.} & F-Value           & Signif.     \\ \hline
device                    & 150.8174 & 9.17E-57                     & 25.1232     & 2.51E-11    \\
$\rho$                       & 9.9004   & 5.62E-05                     & 8.6049      & 2.00E-04    \\
$\lambda$                    & 535.8116 & 3.60E-151                    & 80.6779     & 8.42E-33    \\
display                   & 31.5610  & 2.62E-08                     & 1.4799      & 2.24E-01    \\ \hline
device : $\rho$                & 0.7838   & 5.36E-01                     & 0.5371      & 7.09E-01    \\
device : $\lambda$             & 18.2971  & 1.98E-14                     & 5.6673      & 1.67E-04    \\
$\rho$ : $\lambda$                & 10.0737  & 5.72E-08                     & 0.5573      & 6.94E-01    \\
device : display            & 0.3742   & 6.88E-01                     & 1.3682      & 2.55E-01    \\
$\rho$ : display               & 0.2668   & 7.66E-01                     & 0.5586      & 5.72E-01    \\
$\lambda$ : display            & 1.9499   & 1.43E-01                     & 5.1031      & 6.27E-03    \\ \hline
device : $\rho$ : $\lambda$         & 1.3828   & 2.00E-01                     & 2.4198      & 1.38E-02    \\
device : $\rho$ : display        & 0.3410   & 8.50E-01                     & 0.3107      & 8.71E-01    \\
device : $\lambda$ : display     & 0.3956   & 8.12E-01                     & 0.1496      & 9.63E-01    \\
$\rho$ : $\lambda$ : display        & 0.7717   & 5.44E-01                     & 0.6527      & 6.25E-01    \\ \hline
device : $\rho$ : $\lambda$ : disp & 0.4598   & 8.84E-01                     & 0.6592      & 7.28E-01    \\ \hline
\end{tabular}
\caption{Letter recognition task: F-value table for Aligned Rank Transform (ART) ANOVA. Values were calculated with the ARTool software package. ``device'' refers to the three simulated electrode grids, while ``display'' refers to the use of an \ac{HMD} or monitor. }
\label{tab:letter-anova}
\end{table}

\begin{figure*}[t]
    \centering
    \includegraphics[width=\linewidth,keepaspectratio]{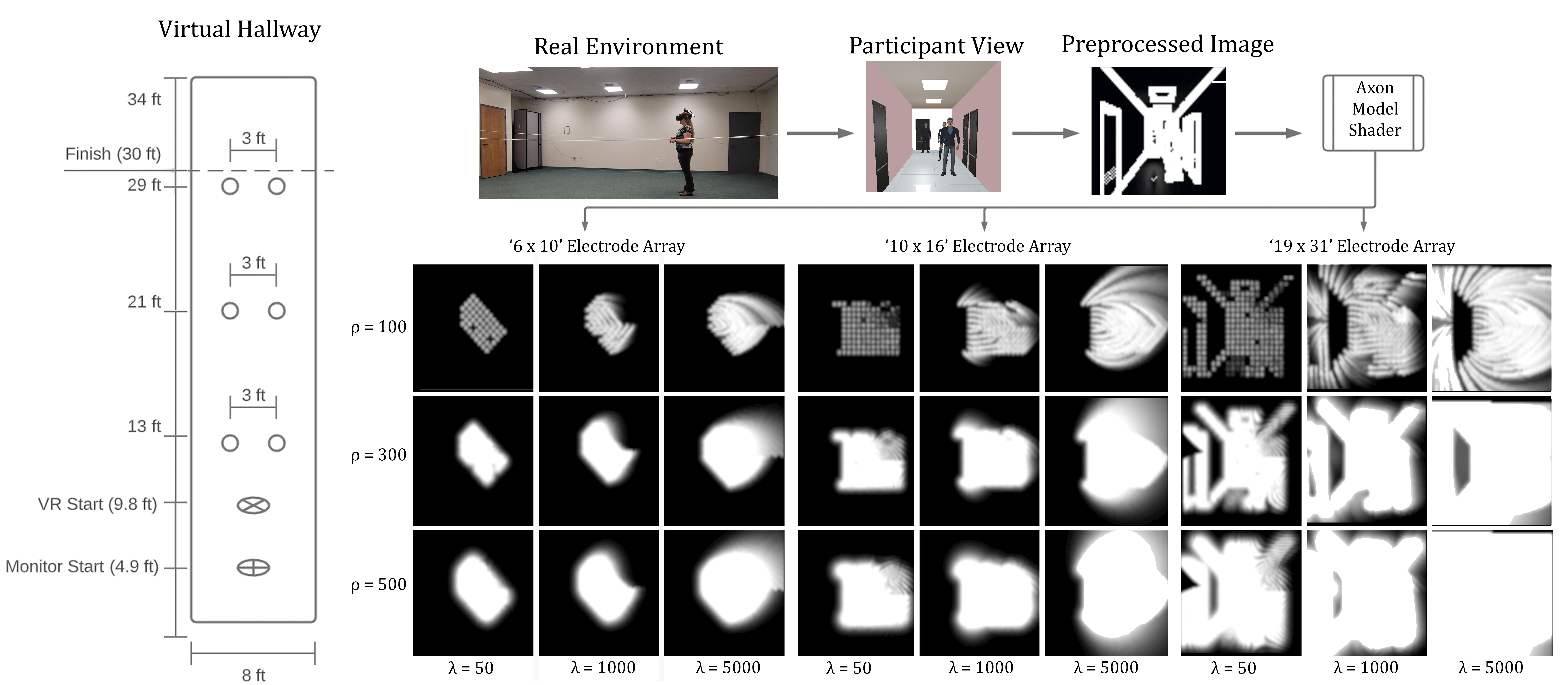}
    \caption{Obstacle avoidance task. \textbf{\emph{Left:}} Layout of the virtual hallway environment modeled after \cite{he_improved_2020}. Empty circles represent the possible locations for obstacles.  \textbf{\emph{Right/Top:}} View of the real environment -> participant's view is passed to the preprocessing shader which performs edge extraction/enhancement before the axon model shader renders SPV. \textbf{\emph{Bottom:}} Output of the axon model shader across the various devices and $\rho$ / $\lambda$ combinations.}
    \Description{Left side is an overhead view of a hallway map.  At the bottom are two locations labelled Monitor Start(4.9 ft) and VR Start(9.8ft). Other positions are labelled further up on the hallway map at 13 feet, 21 feet, and 29 feet.  At each of these locations, there are two circles on either side of the hallway labelled as 3 feet apart. A finish line is marked at 30 feet and the entire hallway is labelled as 8 feet by 34 feet.  Right side is a picture of room with two ropes and a person standing between them with a Virtual Reality headset (labeled ``Real Environment''). An arrow points to the next panel which has a virtual version of a hallway with three people standing in it (labeled ``Participant View''). An arrow points to a black and white rendering of the image where borders between people, walls, and doors have been thickened (labeled ``Preprocessed Image'').  An arrow points to a square labeled ``Axon Model Shader'' that points to three panels displaying simulated prosthetic vision representation of the hallway across three devices }
    \label{fig:exp-hallway-navigation}
\end{figure*}

As expected, increasing the number of electrodes (Fig.~\ref{fig:exp-letter-recognition}A) significantly increased F1 scores in both \ac{HMD} (light gray) and monitor (dark gray) versions of the task.
It is worth noting that participants were consistently above chance levels, even with the simulated Argus II ($6 \times 10$ electrodes) device.
Increasing the number of electrodes also decreased the time it took participants to identify the letter (Fig.~\ref{fig:exp-letter-recognition}D).
However, increasing the number of electrodes from $10 \times 16$ to $19 \times 31$ did not further decrease recognition time.

Contrary to previous findings, F1 scores and recognition time did not systematically vary as a function of phosphene size ($\rho$, Fig.~\ref{fig:exp-letter-recognition}B, E).
In both \ac{HMD} and monitor-based conditions, median F1 scores were highest for $\rho=\SI{300}{\micro\meter}$ (Table \ref{tab:average-letter}).
However, participants achieved similar scores with  $\rho=\SI{100}{\micro\meter}$ in the HMD version and with  $\rho=\SI{500}{\micro\meter}$ in the monitor-based version of the task.

The most apparent differences in performance were found as a function of phosphene elongation ($\lambda$, Fig.~\ref{fig:exp-letter-recognition}C, F).
Using $\lambda=\SI{50}{\micro\meter}$, participants achieved a perfect median F1 score of $1.0$, but this score dropped to $0.741$ for $\lambda=\SI{1000}{\micro\meter}$ and $0.185$ for $\lambda=\SI{5000}{\micro\meter}$ (Table \ref{tab:average-letter}).
Increasing $\lambda$ also significantly increased the time it took participants to identify the letter.

A trend toward a higher F1 score when using the \ac{HMD} was observed across all conditions (Fig.~\ref{fig:results-letter}, \emph{Top}), but the trend failed to reach significance for the device with the lowest number of electrodes ($6 \times 10$ array) or across the larger distortion parameters ($\rho$=\SI{1000}{\micro\meter} and $\lambda$=\SI{5000}{\micro\meter}) (Fig.~\ref{fig:results-letter}, \emph{Top}). 
While average time per trial was faster across all conditions with the HMD, the effect was not significant (Fig.~\ref{fig:exp-letter-recognition}, \emph{Bottom}).

\subsection{Task 2: Obstacle Avoidance}

\subsubsection{Original Task}

The second task was modeled after an obstacle avoidance experiment performed by Argus II recipients~\cite{he_improved_2020}.
In this task, participants were required to walk down a crowded hallway with one to three people located at one of four fixed distances on either the left or right side of the hallway. Participants were permitted the use of a cane and were allowed to touch the walls with the cane (but not the standing persons). Participants were given unlimited time to complete the task and were closely monitored by the experimenter to avoid dangerous collisions. For each trial, the experimenter instructed the participant to stop when they reached the end of the hallway.

\subsubsection{Experimental Setup and Procedure} 

To emulate the experiment described in \cite{he_improved_2020}, we designed a virtual hallway (Fig.~\ref{fig:exp-hallway-navigation}, \emph{Left}) modeled closely after the description and pictures of the physical hallway.

Participants were tasked with successfully navigating the virtual hallway while avoiding collisions with obstacles (simulated people). Each trial consisted of navigating past either two or three obstacles (three trials per condition, six trials total) located on either the left or right side of the hallway (Fig.~\ref{fig:exp-hallway-navigation}). 

\begin{figure*}[t]
    \centering
    \includegraphics[width=\textwidth]{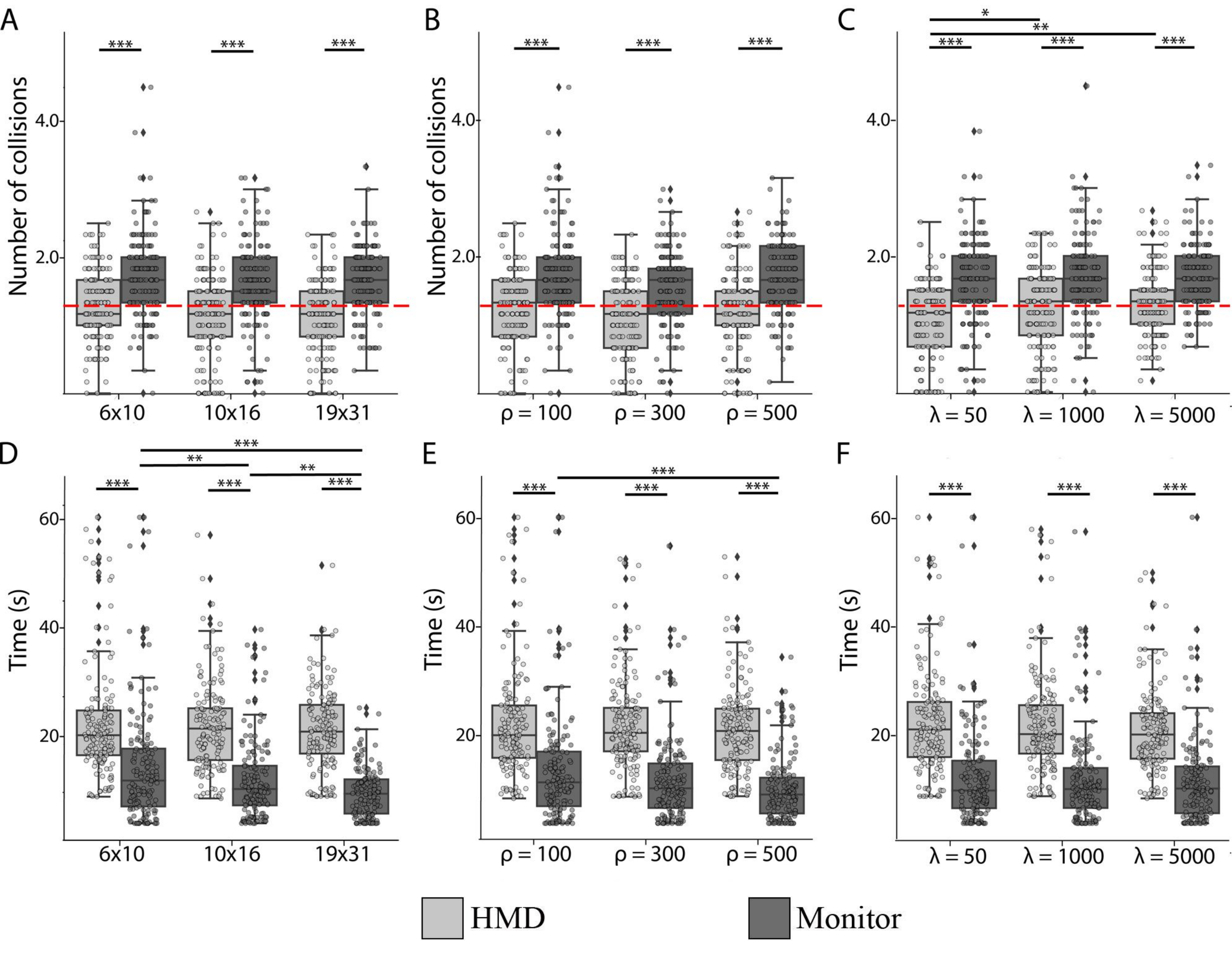}
    \caption{Obstacle avoidance. Data points represent each subject's average performance in a block with boxplots displaying median and interquartile ranges.  \textbf{\emph{Top: }}  Average number of collisions across blocks for each subject within the condition specified by the x-axis. Red line represents chance level (1.25 collisions).
    \textbf{\emph{Bottom}}: Average time across blocks for each subject within the condition specified by the x-axis. Statistical significance was determined using ART ANOVA (*<.05, **<.01, ***<.001).}
    \Description{Figure 6: Six box plots for the hallway task that are fully explained plot by plot in the results section}
    \label{fig:results-hallway}
\end{figure*}

To acclimate participants to the task and controls, we had them perform three initial practice rounds using normal vision. After that, participants completed three more practice rounds with a high-resolution scoreboard model ($31 \times 19$ electrodes, $\rho = \SI{100}{\micro\meter}$, $\lambda = \SI{50}{\micro\meter}$).
Participants were instructed to complete the trials as quickly as possible while avoiding collisions. They were informed that collisions would result in audio feedback; a sample of each sound was played at the beginning of the experiment.

Each combination of 3 devices $\times$ 3 $\rho$ values $\times$ 3 $\lambda$ values were implemented as a block, resulting in a total of 27 blocks. Block order was randomized and participants completed six trials per block for a total of 162 trials for each version (\ac{HMD}/monitor) of the task.
Participants were limited to 1 minute per trial, after which vision was returned to normal and participants walked to the end of the hallway to begin the next trial.

To ensure the safety of participants during the HMD-based version of the task, we positioned rope at the real-life location corresponding to each wall of the hallway (Fig~\ref{fig:exp-hallway-navigation}, \emph{Top, Left}). 
The rope served to guide the participants safely along the path while keeping them in bounds, but was also a substitution for the cane usage in the previous research.
This substitution was necessary, because our testing facility was much larger than the hallway in the original experiment; thus the virtual walls did not coincide with physical walls.

An experimenter was always nearby to ensure the safety of the participants but did not otherwise interact with them during the experiment.
At the end of each trial, the screen turned red and on-screen text instructed participants to turn around and begin the next trial in the other direction.

The monitor version of the task was similar, but each new trial would start automatically without the subject needing to turn around. 
Participants were seated in front of a monitor and were able to use the keyboard to move and the mouse to look around.
The size of the hallway and positions of the obstacles were identical between versions, but participants started 1.5m closer to the first obstacle in the \ac{HMD} version due to size restrictions of the room.

Collisions were detected using Unity's standard continuous collision detection software, with each obstacle having a $\SI{0.7}{\meter} \times \SI{0.4}{\meter}$ hitbox and the participant having a radius of \SI{0.4}{\meter}. Subject locations and orientations were continuously recorded. 
Time per trial, along with individual positions and timings of each collision, were recorded for each trial.  

\begin{table*}[t]
	\centering
	\begin{tabular}{|c|cc|cc|}
		\hline
		& \multicolumn{2}{c|}{Number of Collisions ($\pm$ Std Dev)} & \multicolumn{2}{c|}{Mean Time (s)  ($\pm$ Std Dev)} \\
 		 & HMD & Monitor & HMD & Monitor  \\
	    \hline
		 06x10 Array  &  1.279 ($\pm$ 0.515 )  &   1.734 ($\pm$ 0.621 ) &  22.138  ($\pm$ 10.449 )  &   14.045  ($\pm$ 10.726 )   \\
		 10x16 Array  &  1.148 ($\pm$ 0.602 )  &   \textbf{1.603} ($\pm$ 0.593 ) &  21.483  ($\pm$ 8.216 )  &   11.622  ($\pm$ 7.274 )   \\
		 19x31 Array  &  \textbf{1.117} ($\pm$ 0.562 )  &   1.663 ($\pm$ 0.498 ) &  \textbf{21.052}  ($\pm$ 7.388 )  &   \textbf{9.234}  ($\pm$ 4.469 )  \\
		\hline
		 $\rho$=100  &  1.253 ($\pm$ 0.536 )  &   1.739 ($\pm$ 0.634 ) &  22.000  ($\pm$ 9.950 )  &   13.378  ($\pm$ 10.025 )  \\
		 $\rho$=300  &  \textbf{1.083} ($\pm$ 0.558 )  &   \textbf{1.553} ($\pm$ 0.546 )  &  21.699  ($\pm$ 8.456 )  &   11.594  ($\pm$ 7.725 )  \\
		 $\rho$=500  &  1.208 ($\pm$ 0.586 )  &   1.709 ($\pm$ 0.523 )   &  \textbf{20.974}  ($\pm$ 7.797 )  &   \textbf{9.929}  ($\pm$ 5.780 )   \\
		\hline
		 $\lambda$=50  &  \textbf{1.037} ($\pm$ 0.573 )  &   \textbf{1.627} ($\pm$ 0.637 )  &  22.233  ($\pm$ 9.523 )  &   12.026  ($\pm$ 8.412 )   \\
		 $\lambda$=1000  &  1.209 ($\pm$ 0.610 )  &   1.686 ($\pm$ 0.607 ) &   21.711  ($\pm$ 9.004 )  &   11.495  ($\pm$ 8.183 )  \\
		 $\lambda$=5000  &  1.297 ($\pm$ 0.473 )  &   1.687 ($\pm$ 0.466 )  &  \textbf{20.728}  ($\pm$ 7.677 )  &   \textbf{11.379}  ($\pm$ 7.850 )    \\
		\hline
	\end{tabular}
	\caption{Obstacle avoidance task: Average performance and time per trial across conditions. Best performances (lowest number of collisions/lowest time) for each grouping are presented
    in bold.}
	\label{tab:average-hallway}
\end{table*}

\subsubsection{Evaluating Performance}

Performance was assessed by counting the number of collisions per trial and the amount of time to complete a trial, with a lower number of collisions or lower time per trial expected on easier trials. Analogous to the first task, these two metrics were averaged across trials in a block for each subject and analyzed using ART ANOVA. Post-hoc analyses were performed on significant groups using the Tukey method for multiple comparison adjustments.

\subsubsection{Results}

Results are summarized in Table~\ref{tab:average-hallway} and Fig.~\ref{fig:results-hallway}.
Each data point in Fig.~\ref{fig:results-hallway} represents a subject's number of collisions (Fig.~\ref{fig:results-hallway}, \emph{Top}) and time to completion (Fig.~\ref{fig:results-hallway}, \emph{Bottom}) averaged across repetitions in a block. Group F-values, along with their significance, are reported in Table \ref{tab: hallway-anova}.

\begin{table}[b]
\begin{tabular}{|l|cc|cc|}
\hline
\multicolumn{1}{|l|}{}    & \multicolumn{2}{c|}{{\bf Num Collisions}}                 & \multicolumn{2}{c|}{{\bf Time}} \\
\multicolumn{1}{|l|}{}    & \multicolumn{1}{c}{F} & \multicolumn{1}{c|}{Signif.} & F         & Signif.  \\ \hline
device                    & 4.7538                & 8.85E-03                          & 7.2265    & 2.51E-11      \\
$\rho$                       & 9.2904                & 1.02E-04                          & 25.1790   & 2.00E-04      \\
$\lambda$                    & 4.8301                & 8.21E-03                          & 19.8199   & 8.42E-33      \\
display                   & 207.3125              & 3.27E-42                          & 335.6442  & 2.24E-01      \\ \hline
device : $\rho$                & 1.2885                & 2.73E-01                          & 3.6222    & 7.09E-01      \\
device : $\lambda$             & 1.2039                & 3.08E-01                          & 3.5733    & 1.67E-04      \\
$\rho$ : $\lambda$                & 0.2015                & 9.38E-01                          & 1.1654    & 6.94E-01      \\
device : display            & 1.3595                & 2.57E-01                          & 6.5119    & 2.55E-01      \\
$\rho$ : display               & 0.3381                & 7.13E-01                          & 0.9380    & 5.72E-01      \\
$\lambda$ : display            & 3.8149                & 2.24E-02                          & 9.0722    & 6.27E-03      \\ \hline
device : $\rho$ : $\lambda$         & 1.0423                & 4.02E-01                          & 3.1542    & 1.38E-02      \\
device : $\rho$ : display        & 0.9071                & 4.59E-01                          & 1.9217    & 8.71E-01      \\
device : $\lambda$ : display     & 0.6814                & 6.05E-01                          & 1.2380    & 9.63E-01      \\
$\rho$ : $\lambda$ : display        & 1.5045                & 1.99E-01                          & 2.0618    & 6.25E-01      \\ \hline
device : $\rho$ : $\lambda$ : disp & 0.9815                & 4.49E-01                          & 2.2511    & 7.28E-01      \\ \hline
\end{tabular}
\caption{Obstacle avoidance task: F-value table for Aligned Rank Transform (ART) ANOVA. Values were calculated with the ARTool software package. ``device'' refers to the three simulated electrode grids, while ``display''refers to the use of an \ac{HMD} or monitor. }
\label{tab: hallway-anova}
\end{table}

Contrary to our expectations, neither the number of electrodes (Fig.~\ref{fig:results-hallway}A) nor phosphene size (Fig.~\ref{fig:results-hallway}B) had a significant effect on the number of collisions.
Although the number of collisions decreased slightly with higher electrode counts (Table~\ref{tab:average-hallway}), this did not reach statistical significance.
The only statistical differences could be found between the scoreboard model ($\lambda$=\SI{50}{\micro\meter}) and axon map models ($\lambda$=\{100, 300\}\SI{}{\micro\meter}) for the HMD-based version of the task. 
However, participants performed around chance levels in all tested conditions.

The time analysis revealed a downward trend in time (better performance) with higher electrode counts, but only among the groupings in the monitor version. This trend in time reached significance for all comparisons within the monitor version (Fig. \ref{fig:results-hallway}D). 
Similarly to comparisons across groupings of $\rho$ values, there was a slight downward trend across the median time taken as phosphene distortion increased (Fig. \ref{fig:results-hallway}, \emph{F}). 

A comparison between the two different versions of the task showed a clear difference in performance, with  performance for the HMD version being drastically higher than the monitor version of the task. This trend reached significance across any grouping of device, $\rho$, or $\lambda$ (Fig.~\ref{fig:results-hallway}, \emph{Top}). There was also a difference in time taken between the versions of the task, with the \ac{HMD} version taking longer for all groupings (Fig.~\ref{fig:results-hallway}, \emph{Bottom}).

\section{Discussion}

\subsection{Using an HMD May Benefit Behavioral Performance}

The present study provides the first side-by-side comparison between \ac{HMD} and monitor versions of different behavioral tasks using \ac{SPV}.
Importantly, we used a psychophysically validated \ac{SPV} model to explore the expected behavioral performance of bionic eye users, for current as well as potential near-future devices, and found that participants performed significantly better in the \ac{HMD} version than the monitor version for both tasks.

In the letter recognition task, participants achieved a higher mean F1 score across all conditions (Table~\ref{tab:average-letter}).
However, this trend was only significant for the hypothetical future devices and smaller phosphene sizes and elongations (Fig.~\ref{fig:results-letter}, \emph{Top}).
While average time per trial was faster across all conditions with the HMD, the effect was not significant (Fig.~\ref{fig:exp-letter-recognition}, \emph{Bottom}).

The difference in performance was even more evident in the obstacle avoidance task, where performance (as measured by number of collisions) for the \ac{HMD} version was significantly higher than the monitor version across all conditions (Fig.~\ref{fig:results-hallway}, \emph{Top}).
It is also worth pointing out that participants were able to complete the task faster with higher electrode counts in the monitor-based version of the task. Since the walking speed was fixed across all conditions, this likely indicates that the task was easier with higher electrode counts.

Overall these results suggest that participants were able to benefit from vestibular and proprioceptive cues provided by head movements and locomotion during the \ac{HMD} version of the task, which is something that is available to real bionic eye users but cannot be replicated by a mouse and keyboard.

\subsection{Increased Phosphene Elongation May Impede Performance}

Whereas previous studies treated phosphenes as small, discrete light sources, here we systematically evaluated perceptual performance across
a wide range of common phosphene sizes ($\rho$) and elongations ($\lambda$).
As expected, participants performed best when phosphenes were circular (scoreboard model: $\lambda = \SI{50}{\micro\meter}$; Tables~\ref{tab:average-letter} and \ref{tab:average-hallway}), and increasing phosphene elongation ($\lambda$) negatively affected performance.

In the letter recognition task, participants using the scoreboard model ($\lambda$=\SI{50}{\micro\meter}) achieved a perfect median F1 score of 1.0 (Fig.~\ref{fig:results-letter}C), which is much better than the behavioral metrics reported with real Argus II patients~\cite{cruz_argus_2013}.
Conversely, performance approached chance levels when increasing $\lambda$ to \SI{5000}{\micro\meter}. 

In the obstacle avoidance task, the only significant findings within one version of the experiment were between the scoreboard model ($\lambda = \SI{50}{\micro\meter}$) and either of the larger $\lambda$ values.
This suggests that elongated phosphenes make obstacle avoidance more challenging than the scoreboard model.
However, participants performed around chance levels in all tested conditions, which was also true for real Argus II patients \cite{he_improved_2020}.

Contrary to our expectations, phosphene size ($\rho$) did not systematically affect performance (Fig.~\ref{fig:results-letter}B, Fig.~\ref{fig:results-hallway}B).
The best performance was typically achieved with $\rho = \SI{300}{\micro\meter}$.
This is in contrast to previous literature suggesting smaller phosphene size is directly correlated with higher visual acuity \cite{chen_simulating_2009, han_deep_2021}

Overall these findings suggest that behavioral performance may vary drastically depending on the choices of $\rho$ and $\lambda$.
This is important for predicting visual outcomes, because  $\rho$ and $\lambda$ have been shown to vary drastically across bionic eye users \cite{beyeler_model_2019}, suggesting future work should seek to use psychophysically validated \ac{SPV} models when making theoretical predictions  about device performance.

\subsection{Increasing the Number of Electrodes Does Not Necessarily Improve Performance}

As expected, letter recognition performance improved as the size of the electrode grid (and therefore the \ac{FOV}) was increased from $6 \times 10$ to $10 \times 16$ and $19 \times 31$ (Fig.~\ref{fig:results-letter}A).
This performance benefit was also observed in the time it took participants to recognize the letter (Fig.~\ref{fig:results-letter}D), and is consistent with previous literature on object recognition \cite{thorn_virtual_2020}.

However, electrode count did not affect behavioral performance in the obstacle avoidance task. 
Whereas there was a slight increase in performance scores for devices with more electrodes (Fig.~\ref{fig:results-hallway}A), this effect 
did not reach significance. 

Overall these results are consistent with previous literature suggesting that, for most tasks, the number of electrodes may not be the limiting factor in retinal implants  \cite{beyeler_learning_2017, behrend_resolution_2011}.

\subsection{Limitations and Future Work}

Although the present study addressed previously unanswered questions about \ac{SPV}, there are a number of limitations that should be addressed in future work as outlined below.

First, in an effort to focus on the impact of phosphene size and elongation on perceptual performance, we limited ourselves to modeling spatial distortions. However, retinal implants are known for causing temporal distortions as well, such as flicker and fading, which may further limit the perceptual performance of participants \cite{beyeler_learning_2017}.

Second, the displayed stimuli were not contingent on the user's eye movements.
Even though current retinal implants ignore eye movements as well, there is a not-so-subtle difference between a real retinal implant and a simulated one. Since the real device is implanted on the retinal surface, it will always stimulate the same neurons, and thus produce vision in the same location in the visual field---no matter the eye position. This can be very disorienting for a real patient as shifting your gaze to the left would not shift the vision generated by the implant. 
In contrast, a participant in a \ac{VR} study is free to explore the presented visual stimuli with their gaze, thus artificially increasing the \ac{FOV} from that offered by the simulated device.
Consequently, the here presented performance predictions may still be too optimistic.
In the future, simulations should make use of eye tracking technologies to update the scene in a gaze-contingent way.

Third, we did not explicitly measure the level of immersion across the two display types (\ac{HMD} and monitor).
Instead, we assumed that viewing a scene that updates with the user's head movement through an \ac{HMD} would lead to a higher level of immersion.
Although this may be true for realistic virtual environments \cite{miller_level_2016}, this has yet to be demonstrated for \ac{SPV} studies.
Future \ac{SPV} work should therefore explicitly measure the level of immersion and/or a user's sense of presence.

Fourth, the obstacle avoidance task did not have a meaningful time metric. Although participants performed the task significantly faster in the monitor-based version, this is likely an artifact due to the walking speed of participants not being consistent between versions of the task. 
Participants moved much slower with the \ac{HMD} as they were not able to see the real world around them. Future studies should take this into consideration and correct for each participant's walking speed within desktop versions of tasks.

Fifth, the study was performed on sighted graduate students readily available at the University of California, Santa Barbara.
Their age, navigational affordances, and experience with low vision may therefore be drastically different from real bionic eye users, who tend to not only be older and prolific cane users but also receive extensive vision rehabilitation training.

Interestingly, we found vast individual differences across the two tasks (individual data points in Figs.~\ref{fig:results-letter} and \ref{fig:results-hallway}) which were not unlike those reported in the literature \cite{cruz_argus_2013,he_improved_2020}.
Subjects who did well in one experiment tended to do well across all versions of both experiments (data not shown), suggesting that some people were inherently better at adapting to prosthetic vision than others.
Future work should therefore zero in on the possible causes of these individual differences and compare them to real bionic eye users.
Studying these differences could identify training protocols to enhance the ability of all device users.

\section{Conclusions}

The present work constitutes a first essential step towards immersive VR simulations of bionic vision. 
Data from two behavioral experiments demonstrate the importance of choosing an appropriate level of immersion and phosphene model complexity.
The VR-SPV toolbox that enabled these experiments is freely available at \url{https://github.com/bionicvisionlab/BionicVisionXR} and designed to be extendable to a variety of bionic eye technologies.
Overall this work has the potential to further our understanding of the qualitative experience associated with different bionic eye technologies and provide realistic expectations of prosthetic performance.

\begin{acks}
This work was supported by the National Institutes of Health (NIH R00 EY-029329 to MB).
\end{acks}

\bibliographystyle{ACM-Reference-Format}
\bibliography{bibliography.bib}

\end{document}